\documentclass[preprint,12pt]{elsarticle}

\usepackage{hyperref}
\usepackage{lineno}
\usepackage{amsmath}
\usepackage{amsfonts}
\usepackage{amssymb}
\usepackage{graphicx}
\usepackage{siunitx}
\usepackage{xcolor}
\usepackage{caption}
\usepackage{subcaption}
\usepackage{booktabs}
\usepackage{graphicx}

\DeclareSIUnit\neq{n_{eq}}
\DeclareSIUnit\electrons{e}

\begin{document}

\begin{frontmatter}

\title{Radiation tolerant, thin, passive CMOS sensors read out with the RD53A chip}

\author{Y. Dieter\corref{cor1}}
\ead{dieter@physik.uni-bonn.de}
\cortext[cor1]{Corresponding author}
\author{M. Daas}
\author{J. Dingfelder}
\author{T. Hemperek}
\author{F. H\"ugging}
\author{J. Janssen}
\author{H. Kr\"uger}
\author{D.-L. Pohl}
\author{M. Vogt}
\author{T. Wang}
\author{N. Wermes}
\author{P. Wolf}
\address{University of Bonn, Physikalisches Institut, Nu{\ss}allee 12, 53115 Bonn, Germany}

\begin{abstract}
The radiation hardness of passive CMOS pixel sensors fabricated in \SI{150}{\nano\meter} LFoundry technology is investigated. CMOS process lines are especially of interest for large-scale silicon detectors as they offer high production throughput at comparatively low cost. Moreover, several features like poly-silicon resistors, MIM-capacitors and several metal layers are available which can help enhance the sensor design. The performance of a \SI{100}{\micro\meter} thin passive CMOS sensor with a pixel pitch of \SI{50}{\micro\meter} at different irradiation levels, \SI{5e15}{\neq\per\square\centi\meter} and \SI{1e16}{\neq\per\square\centi\meter}, is presented. The sensor was bump-bonded and read out using the RD53A readout chip. After the highest fluence a hit-detection efficiency larger than \SI{99}{\percent} is measured for minimum ionising particles. The measured equivalent noise charge is comparable to conventional planar pixel sensors.
Passive CMOS sensors are thus an attractive option for silicon detectors operating in radiation harsh environments like the upgrades for the LHC experiments.
\end{abstract}


\end{frontmatter}


\section{Introduction}
After the upgrade of the Large Hadron Collider (LHC) at CERN in 2026~\cite{apollinari2017high}, the accelerator aims to deliver \num{5} - \num{7} times the nominal lu\-mi\-no\-sity of \SI{1e34}{\per\square\centi\meter\per\second}.
This results in new challenges in terms of data rate capabilities and radiation tolerance, especially for detector layers close to the interaction point of the colliding proton beams. Therefore, the ATLAS and CMS experiments will replace their currently installed tracking detectors with detectors featuring larger areas of silicon and decreased pixel pitch \cite{CERN-LHCC-2017-021, CERN-LHCC-2017-009}.
In particular, the ATLAS Inner Detector \cite{Aad_2008} is going to be replaced by an all-silicon tracking detector, the ATLAS Inner Tracker (ITk) \cite{CERN-LHCC-2017-021}. The expected \SI{1}{\mega\electronvolt} neutron equivalent fluence\footnote{after an integrated luminosity of \SI{2000}{\per\femto\barn}} for the innermost layer is approximately \SI{1e16}{\neq\per\square\centi\meter}, for the outer layers fluences\footnote{after an integrated luminosity of \SI{4000}{\per\femto\barn}} from \SI{2e15}{\neq\per\square\centi\meter} up to \SI{5e15}{\neq\per\square\centi\meter} are expected \cite{CERN-LHCC-2017-021}.

Since it was demonstrated in the past that hybrid pixel detectors \cite{rossi2006pixel} can be successfully operated in such harsh radiation environments at the LHC, it is foreseen to employ this technology also for the upgraded pixel detector.
However, the surface of the new ATLAS pixel detector is increased from approximately \SI{2}{\square\meter} to \SI{13}{\square\meter} \cite{CERN-LHCC-2017-021}, demanding large-area solutions with cost-effective designs. 
An approach of employing monolithic (active) CMOS pixel detectors \cite{PERIC2007876,WERMES2016483} which combine the sensing and electronic processing functions, and thus sa\-ving the time-consuming and expensive hybridisation process, has been investigated \cite{Wang_2018, Caicedo_2019}.
With this CMOS pixel development an interesting option became attractive that utilises commercial CMOS process lines for the fabrication of planar sensors as the sensing part of hybrid pixel detectors. CMOS fabrications offer a high production throughput at comparatively low cost. Further benefits arise from the fact that several features are available to enhance the sensor design. For instance, polysilicon layers can be used to connect each pixel to a bias grid making it possible to test the functionality of the sensor at wafer level. Also, MIM (metal-insulator-metal) capacitors can be used to AC-couple the sensor pixels to the readout chip pixels preventing leakage current flowing into the readout channels. Moreover, several metal layers are available that can be exploited as re-distribution layers such that enlarged inter-gap pixels (between two readout chips) can be avoided.
These sensors are called \textit{passive} CMOS sensors since they do not have any active components implemented.

Passive CMOS sensors in a large pixel pitch design ($\num{50}\times\SI{250}{\square\micro\meter}$ pixels) were already investigated \cite{pohl_cmos}. The characterisation of the small pixel pitch design ($\num{50}\times\SI{50}{\square\micro\meter}$ pixels) before irradiation can be found in \cite{dieter2020}. 
In the following the performance of irradiated passive CMOS sensors is studied to demonstrate their radiation tolerance and suitability for the upgrades of the LHC experiments.

\section{LFoundry passive CMOS pixel sensor}
\subsection{Pixel sensor design}
\begin{figure}[t]
	\begin{center}
		\includegraphics[width=0.65\linewidth]{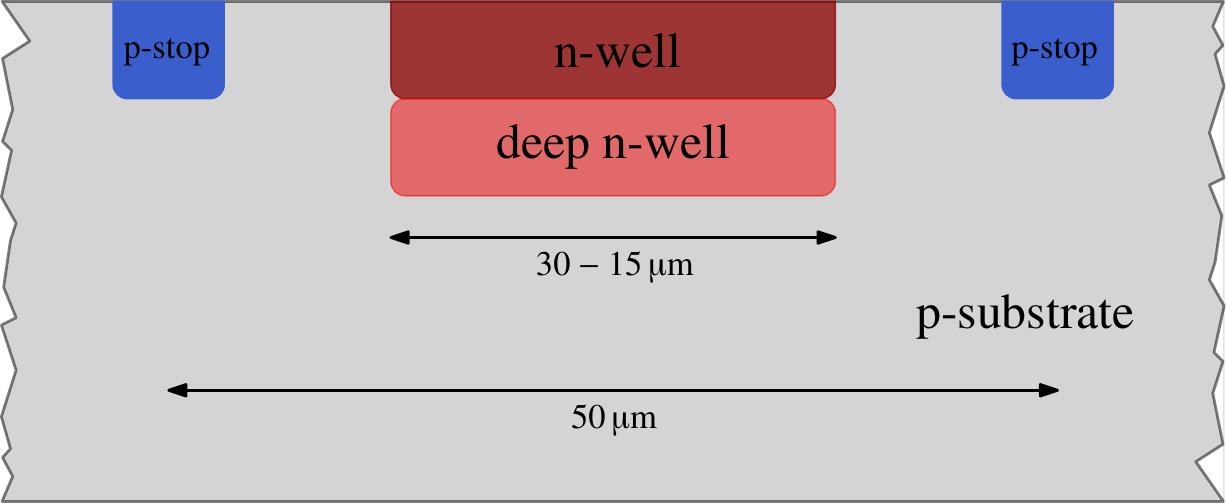}
		\caption{Simplified schematic cross section of an n-in-p pixel of the LFoundry passive CMOS pixel sensor. The charge collection electrode is an n-well (optionally with an additional deep n-well) with varying implant size between \SI{15}{\micro\meter} and \SI{30}{\micro\meter}. The polysilicon layer is omitted. For details see Fig.~\ref{fig:lfcmos_layout}. Small p-implantations (p-stop) isolate the pixels from each other.}
		\label{fig:lfcmos_pixel}
	\end{center}
\end{figure}

Passive CMOS n-in-p pixel sensors in \SI{150}{\nano\meter} LFoundry \cite{lfoundry} technology were manufactured on high-resistivity Czochralski wafers. The resistivity of the substrate is at least \SI{2}{\kilo\ohm\centi\meter}, as specified by the foundry. Measurements suggest that the resistivity is between \num{5} and \SI{6}{\kilo\ohm\centi\meter} \cite{pohl_cmos}. The substrate was thinned to a thickness of \SI{100}{\micro\meter}. The backside was processed by Ion Beam Services (IBS) \cite{IBS} including backside implantation as well a backside me\-ta\-llisation allowing for backside bias application.
The sensor consists of $\num{64} \times \num{64}$ pixels with a size of $\num{50} \times \SI{50}{\square\micro\meter}$, and has a total area of $\num{3.8} \times \SI{3.8}{\square\milli\meter}$.

\begin{figure}[t]
	\begin{center}
		\includegraphics[width=0.78\linewidth]{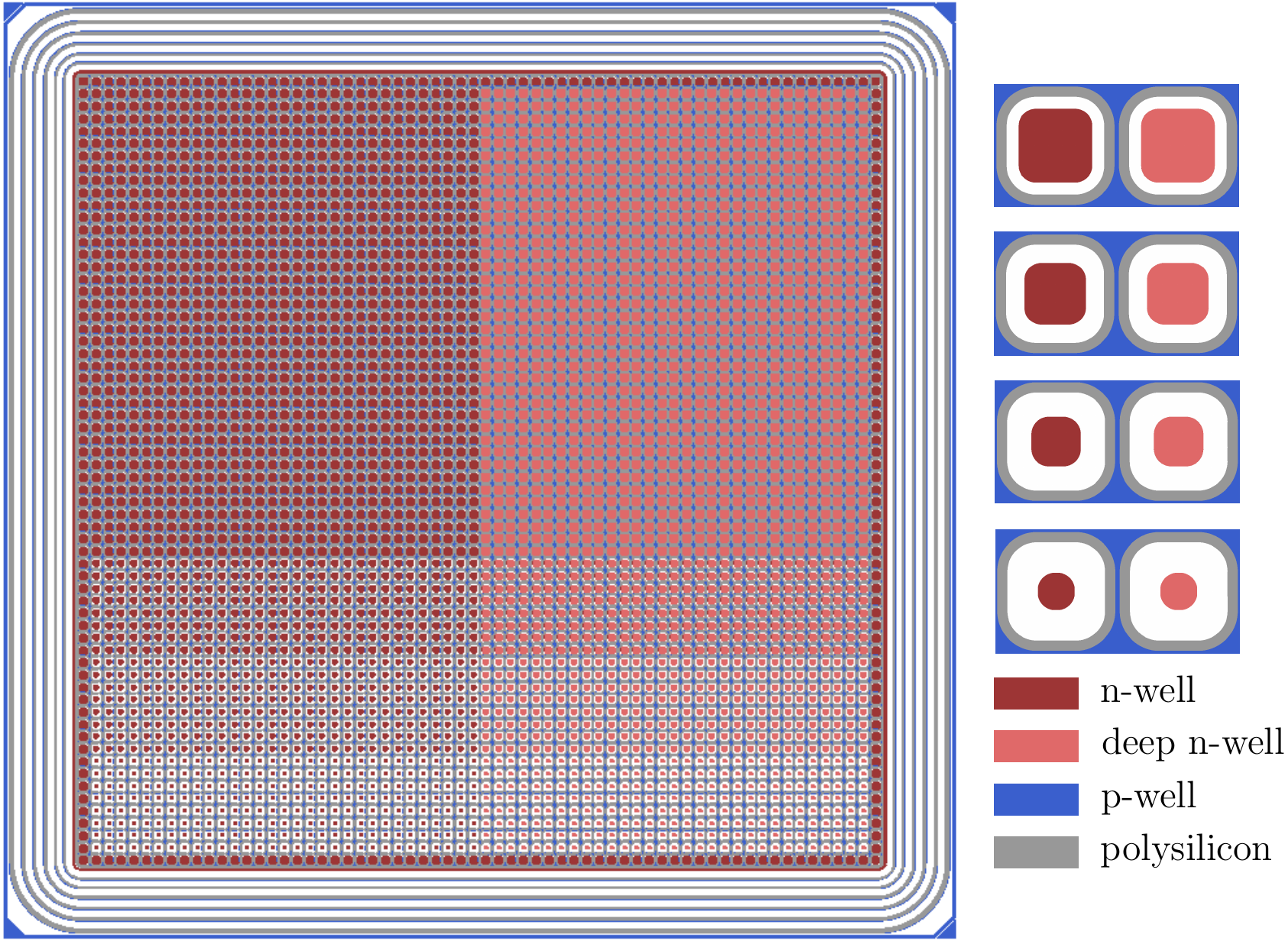}
		\caption{Left: Layout of the LFoundry passive CMOS pixel sensor. The left half of the pixel matrix consists of pixels with a n-implantation (NW), whereas the right half consists of pixels with an additional deep n-implantation (DNW). The size of the n-implantations varies in both dimensions between \SI{30}{\micro\meter} (top of the matrix) and \SI{15}{\micro\meter} (bottom of the matrix). Guard-rings isolate the pixels from the high voltage at the sensor edge. Right: Enlarged view of the different pixel designs.}
		\label{fig:lfcmos_layout}
	\end{center}
\end{figure}

Fig.~\ref{fig:lfcmos_pixel} shows a simplified schematic cross section of one pixel, Fig.~\ref{fig:lfcmos_layout} depicts the layout of the pixel matrix.
In order to investigate the charge collection properties and the pixel capacitance, various pixel designs were implemented. The left half of the pixel matrix consists of pixels with a regular n-implantation (n-well, denoted as NW), whereas the right half of the pixel matrix consists of pixels with an additional deep n-implantation (deep n-well, denoted as DNW). The size of the n-implantations varies in both dimensions between \SI{30}{\micro\meter} (top of the matrix) and \SI{15}{\micro\meter} (bottom of the matrix). To isolate the pixels from each other a small p-implantation (p-stop) is used. Moreover, a fine-pitch polysilicon layer encloses the n-implantations with the intention to increase the breakdown voltage, especially after irradiation.
The pixel matrix is surrounded by an n-implantation confining the active pixel region. In addition, six guard-rings isolate the pixels from the high voltage at the sensor edge.

The sensor is bump-bonded via solder bumps (by Fraunhofer IZM \cite{izm}) to the RD53A readout chip \cite{Garcia-Sciveres:2287593, Monteil:2019niy}, a prototype readout chip for the ATLAS ITk pixel detector.

\subsection{Pixel sensor irradiation}
The studied pixel detector has been step-wise irradiated to the target fluence to investigate the performance at different irradiation levels. In the first step, the detector was irradiated to a fluence of \SI{5e15}{\neq\per\square\centi\meter} at the MC40 cyclotron of the University of Birmingham \cite{Allport_2017} using \SI{27}{\mega\electronvolt} protons. In the second step, the detector was irradiated to a total fluence of \SI{1e16}{\neq\per\square\centi\meter} at the Proton Irradiation Site at the Bonn Isochronous Cyclotron \cite{wolf_ma} using \SI{14}{\mega\electronvolt} protons. The irradiations were performed uniformly in a cold environment and the device was left unpowered during irradiation. After each irradiation step the device was annealed for \SI{80}{\minute} at \SI{60}{\celsius}.
The co\-rres\-ponding total ionising doses\footnote{Important for surface damage affecting the readout chip.} created by protons were estimated to approximately \SI{660}{\mega\radian} (Birmingham) and \SI{580}{\mega\radian} (Bonn).

\section{Leakage current measurements}

\begin{figure}[t]
	\begin{center}
		\includegraphics[width=0.7\linewidth]{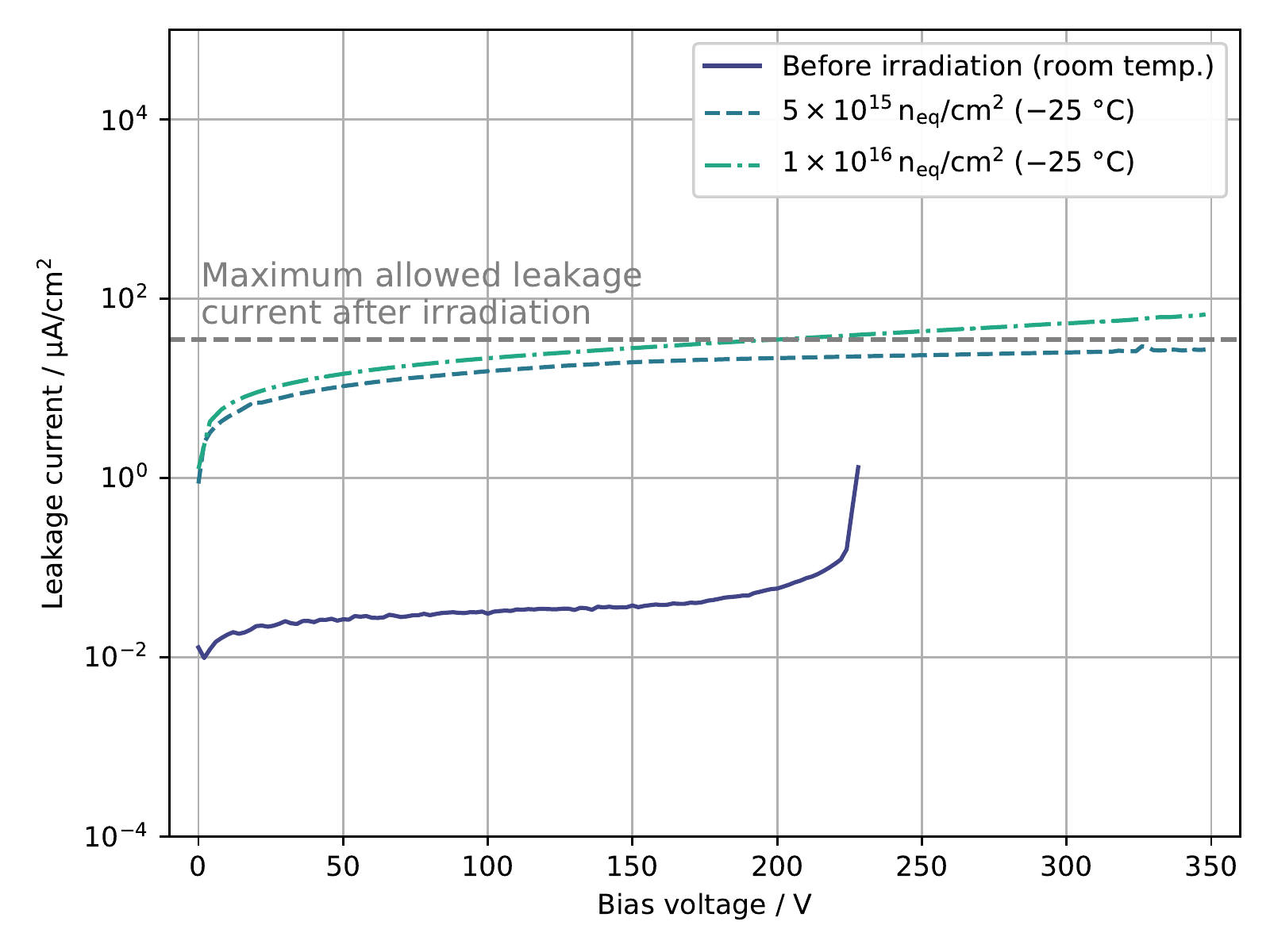}
		\caption{Leakage current as a function of the (reverse) applied bias voltage before (solid line) and after irradiation (dashed lines) to a fluence of \SI{5e15}{\neq\per\square\centi\meter} and \SI{1e16}{\neq\per\square\centi\meter}. The grey dashed line indicates the maximum allowed leakage current of \SI{35}{\micro\ampere\per\square\centi\meter} after a fluence of \SI{5e15}{\neq\per\square\centi\meter} according to the ATLAS ITk specifications. The leakage current is normalised to the total area of the sensor.}
		\label{fig:iv_curve}
	\end{center}
\end{figure}

To test the functionality of the sensors and to determine their maximum operational voltage, the leakage current is measured as a function of the (reverse) bias voltage (IV-curve). Fig.~\ref{fig:iv_curve} shows the IV-curves before and after irradiation of the tested sensor after bump-bonding. The IV-curve before irradiation was measured at room temperature, whereas the IV-curves after irradiation were measured at an environmental temperature of \SI{-25}{\celsius}. Before irradiation the maximum operational voltage is approximately \SI{220}{\volt}. After irradiation the sensor is still functional and there is no breakdown visible anymore up to the maximum tested operational voltage of \SI{350}{\volt}. Furthermore, the leakage current after a fluence of \SI{5e15}{\neq\per\square\centi\meter} is approximately \SI{23}{\micro\ampere\per\square\centi\meter} and meets the requirements for ATLAS ($<\SI{35}{\micro\ampere\per\square\centi\meter}$). As expected, the leakage current at a fluence of \SI{1e16}{\neq\per\square\centi\meter} is approximately twice as large as for a fluence of \SI{5e15}{\neq\per\square\centi\meter}.
The power dissipation of the sensor needed for a hit-detection efficiency larger than \SI{97}{\percent} (see Sec.~\ref{sec:eff}) is about \SI{7}{\milli\watt\per\square\centi\meter} (\SI{35}{\micro\ampere\per\square\centi\meter} at \SI{200}{\volt}) after a fluence of \SI{1e16}{\neq\per\square\centi\meter}. This is comparable to the power dissipation reported for 3D sensors \cite{Terzo:2744099}.

\section{Electronic noise}
\begin{figure}[t]
	\begin{center}
		\includegraphics[width=0.7\linewidth]{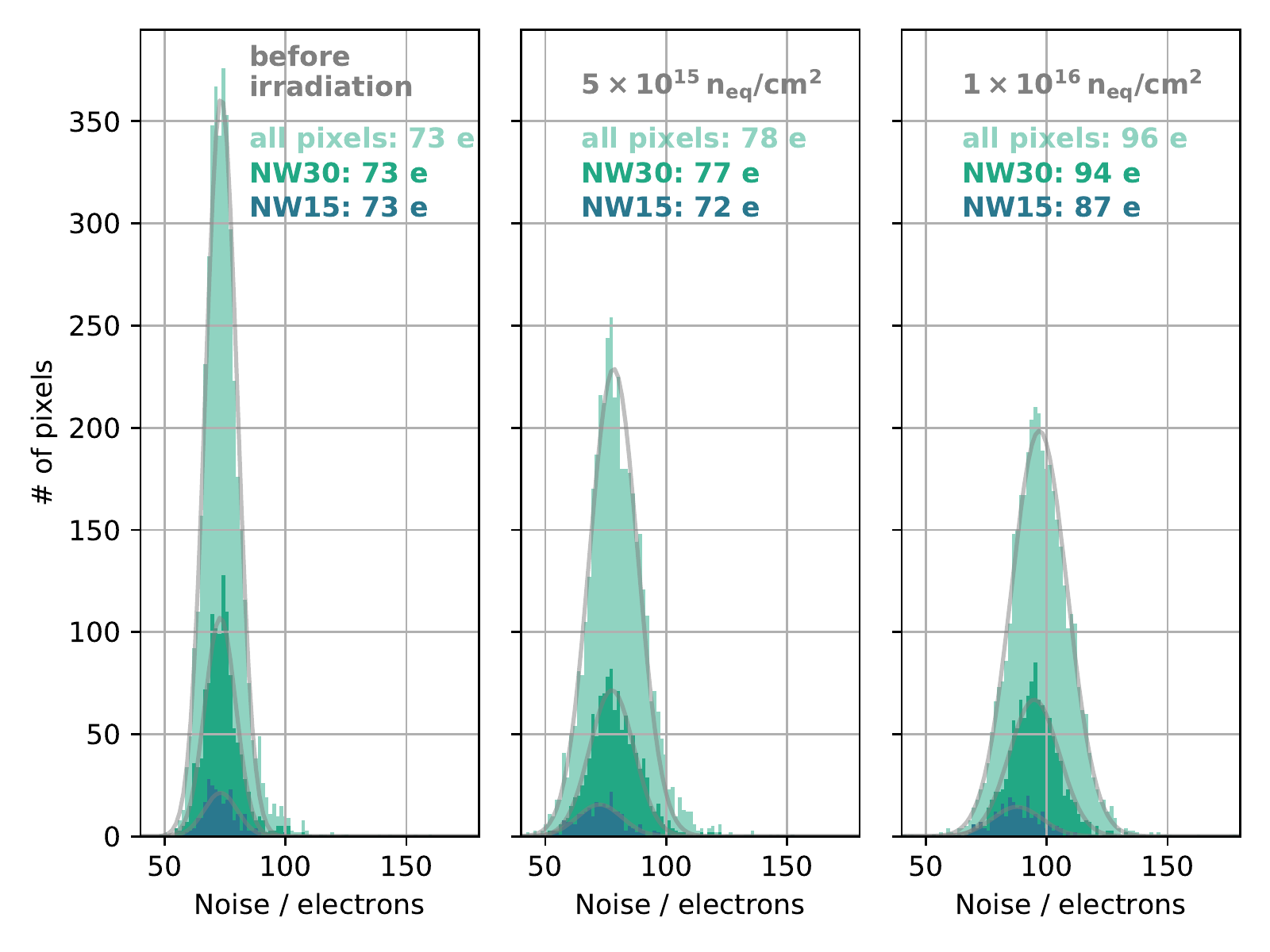}
		\caption{Equivalent noise charge distributions at different irradiation levels. The noise distributions for different pixel geometries are shown exemplary for the NW30 and NW15 pixels. The distributions are fitted with a Gaussian function to extract the mean noise. The uncertainties of the estimated means are approximately \SI{2}{\electrons}.}
		\label{fig:noise_map}
	\end{center}
\end{figure}

An important parameter to quantify the performance of a sensor is the equivalent noise charge (ENC). The ENC distributions of the investigated pixel detector at different irradiation steps can be seen in Fig.~\ref{fig:noise_map}.

Before irradiation, an ENC of \SI{73}{\electrons} is measured. This is a value comparable to other planar sensor designs read out with the same amplifier chip. After irradiation, the ENC increases to about \SI{100}{\electrons}. The reason for that is most likely an increase in shot noise due the higher leakage current after irradiation (approximately \SI{90}{\micro\ampere} corresponding to \SI{22}{\nano\ampere} per pixel\footnote{Measured at a temperature of \SI{-17}{\celsius}.}). In addition, the performance of the analogue front-end is degraded by irradiation (i.e. the transconductance $g_m$ decreases) and is likely responsible for an unspecifiable additional noise contribution. Further, it cannot be excluded that the detector capacitance increases after irradiation which would also lead to an increase in noise.

Comparing different pixel designs, there is no significant difference in terms of noise observed before irradiation, although the measured pixel capacitances\footnote{Including contributions due to routing and bump-bonds.} are different for the various pixel geometries, \SI[separate-uncertainty=true]{33.5(2)}{\femto\farad} for NW30 pixels and \SI[separate-uncertainty=true]{22.4(2)}{\femto\farad} for NW15 pixels \cite{Kr_ger_2021}.
At a fluence of \SI{1e16}{\neq\per\square\centi\meter}, the noise of NW30 pixels is approximately \SI{8}{\percent} higher than for NW15 pixels.
Since this difference is small, the benefits in terms of noise reduction for pixels featuring small implants do not outweigh the disadvantages that arise in terms of hit-detection efficiency (see Sec.~\ref{sec:eff}).

\section{Hit-detection efficiency measurement}\label{sec:eff}
A crucial detector property is the hit-detection efficiency, i.e. the probability with which a hit (a particle traversing a pixel) is recognised by the detector. For the application as a tracking detector the hit-detection efficiency has to be high for efficient hit finding and track reconstruction. Especially after irradiation, the hit-detection efficiency is of interest, and for ATLAS ITk it is required to be above \SI{97}{\percent} \cite{CERN-LHCC-2017-021}.

To measure the hit-detection efficiency the device under test (DUT) is placed in a beam-telescope setup consisting of six high-resolution Mimosa26 planes (EUDET-type beam telescope) \cite{Jansen2016} and an ATLAS FE-I4 \cite{GARCIASCIVERES2011S155} re\-fe\-rence plane. The Mimosa26 planes provide a high spatial resolution of approximately \SI{3}{\micro\meter} \cite{Jansen2016} allowing a precise track reconstruction. However, the time resolution of the Mimosa26 sensors is limited due to their rolling shutter readout (duration of \SI{115.2}{\micro\second}). In contrast, the FE-I4 reference plane provides a very good time-stamping capability with a precision better than \SI{25}{\nano\second}. 
The challenge during track reconstruction is to ensure a correct time assignment of the Mimosa26 tracks, which is needed for a proper hit-detection efficiency measurement. Therefore, the ATLAS FE-I4 plane is used as a time reference plane such that the tracking hits in the Mimosa26 planes spatially match the hit in the timing reference plane, which thus provides the time stamp for the track.

The hit-detection efficiency of the investigated sensor was measured using a minimum ionising electron beam provided by the ELSA test beam facility~\cite{Heurich:2016ilc} (\SI{2.5}{\giga\electronvolt}) and the DESY II test beam facility \cite{Diener_2019} (\SI{5}{\giga\electronvolt}).
A scin\-ti\-lla\-tor in front of the telescope setup generates a trigger signal when particles traverse the setup. An EUDET-type Trigger Logic Unit (TLU) \cite{tlu} is used to distribute and synchronise the trigger signals with the different readout systems. The Mimosa26 telescope is read out without being triggered in continuous rolling shutter readout mode using the \textit{pymosa} software~\cite{pymosa}. The DUT is read out triggered using the \textit{BDAQ53} software~\cite{Daas_2021} and the ATLAS FE-I4 plane was read out triggered using the \textit{PyBAR} software~\cite{pybar}. The data is analysed using the \textit{beam telescope analysis} software~\cite{bta} including clustering, detector alignment, and track reconstruction as well as the final result analysis of the hit-detection efficiency and charge collection behaviour.

For all following measurements, the detector was tuned to a threshold of approximately \SI{1000}{\electrons} with a noise occupancy of less than \num{e-6} per pixel.

\begin{figure}
	\begin{center}
		\includegraphics[width=0.75\linewidth]{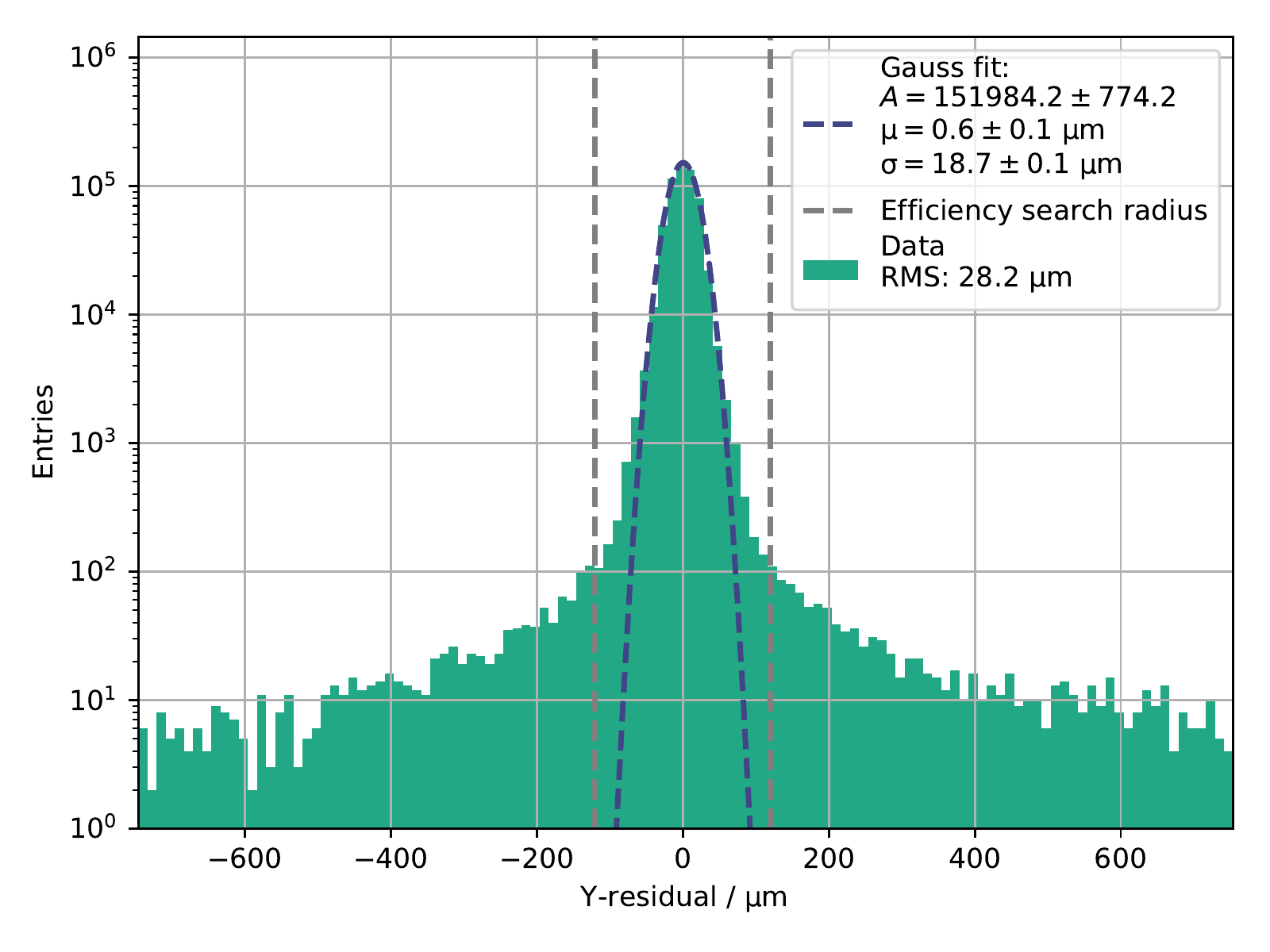}
		\caption{(Unbiased) residual distribution in one dimension at the DUT. The data is shown on a logarithmic scale. The distribution is fitted with a Gaussian function. The grey dashed line illustrates the maximum distance between a hit and track intersection (with the DUT) at which a hit contributes to the efficiency.}
		\label{fig:residuals}
	\end{center}
\end{figure}

Fig.~\ref{fig:residuals} illustrates the (unbiased) residual distribution (distance between hit and track intersection) in the y-dimension at the DUT. The residuals are centred around zero which indicates a correct alignment of the detector planes. Due to multiple scattering the residual distribution is smeared out and can be approximated with a Gaussian function. The deviation from the Gaussian function towards the tails originates from the fact that the probability for large scattering angles is enhanced as described in Molière's theory~\cite{moliere}. From a fit a residual width (1-$\sigma$ width) of \SI{18.7}{\micro\meter} is extracted. This is in agreement with the expectation since the residual width for unbiased tracks is the quadratic sum of the intrinsic resolution of the DUT ($\frac{\mathrm{pixel\,\,pitch}}{\sqrt{12}}$) and the pointing resolution at the DUT (a few \si{\micro\meter}). The pointing resolution in this setup is slightly worsened by the additional material of the cooling infrastructure (cooling box for DUT and PCB cooling plate). However, the resolution is still sufficient for in-pixel studies.

\begin{figure}
\centering
\begin{subfigure}{.5\textwidth}
  \centering
  \includegraphics[width=1.0\linewidth]{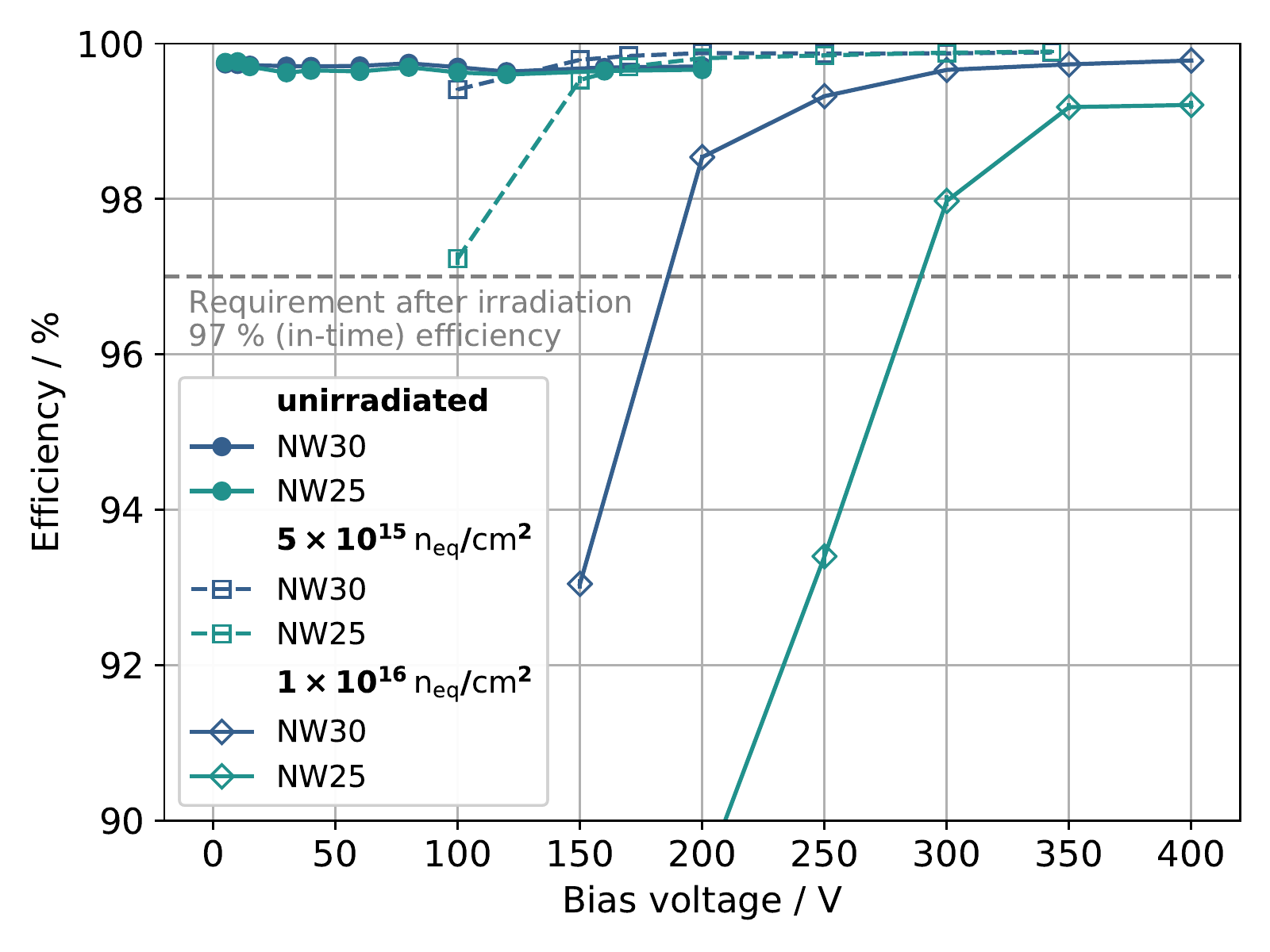}
  \caption{Hit-detection efficiency of NW-flavors}
  \label{fig:eff_vs_bias_nw}
\end{subfigure}%
\begin{subfigure}{.5\textwidth}
  \centering
  \includegraphics[width=1.0\linewidth]{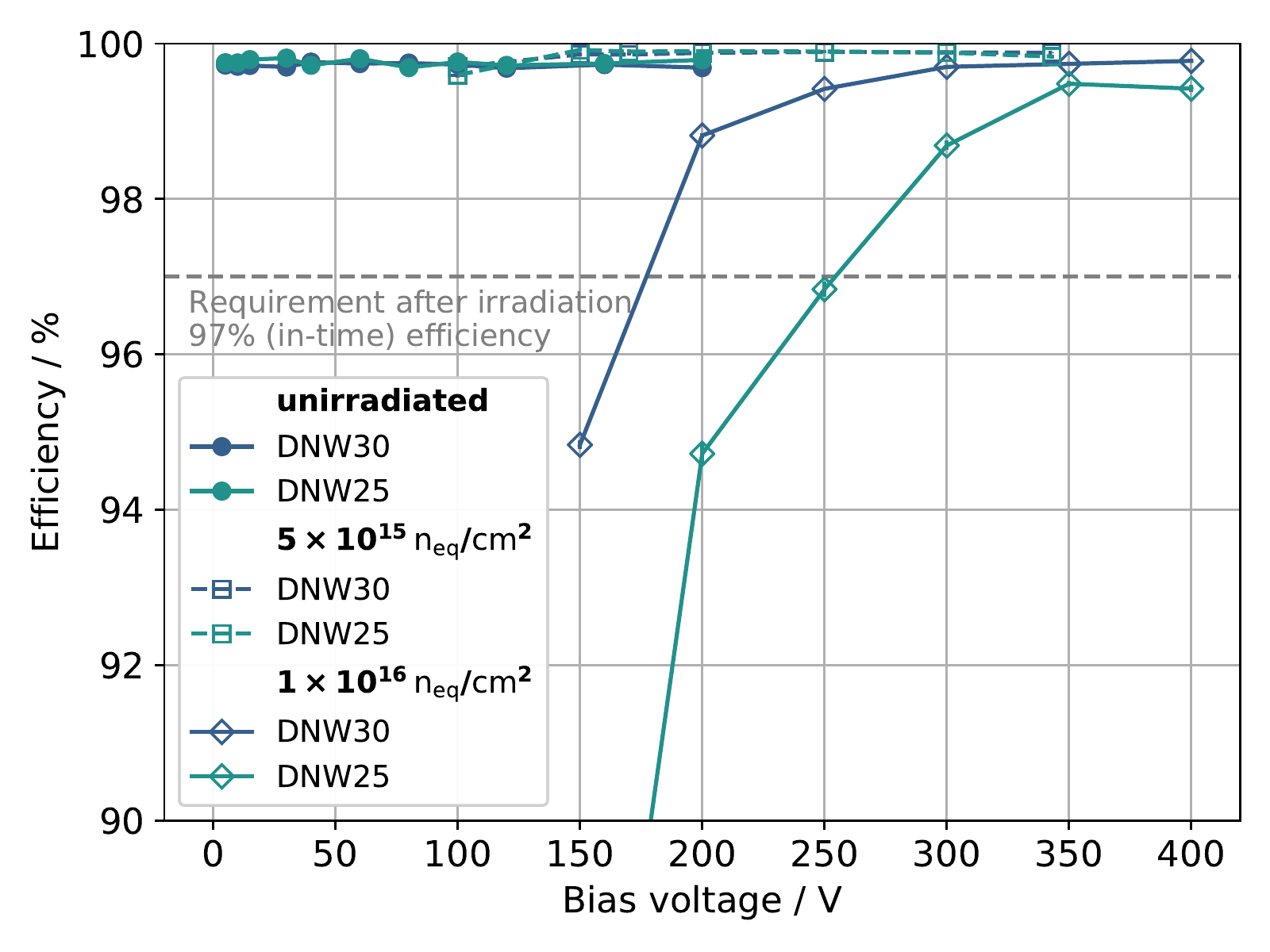}
  \caption{Hit-detection efficiency of DNW-flavors}
  \label{fig:eff_vs_bias_dnw}
\end{subfigure}
\caption{Hit-detection efficiency as a function of bias voltage for different pixel flavors and different irradiation levels. The grey dashed line represents the requirement of an (in-time) efficiency larger than \SI{97}{\percent}. For the sake of clarity, not all pixel designs are shown. The quoted error bars are purely statistical. Left: NW-flavors. Right: DNW-flavors.}
\label{fig:eff_vs_bias}
\end{figure}

In order to reject noise hits (spatially uncorrelated) which artificially increase the hit-detection efficiency, hits are only considered as efficient if the residual of the track is smaller than a given distance (of \SI{120}{\micro\meter}). This efficiency search radius is illustrated in Fig.~\ref{fig:residuals}.

Fig.~\ref{fig:eff_vs_bias_nw} shows the hit-detection efficiency before and after irradiation as a function of the applied (reverse) bias voltage for the NW30 and NW25 pixel designs (regular n-implantation with a size of \num{30} and \SI{25}{\micro\meter}), Fig.~\ref{fig:eff_vs_bias_dnw} shows this for the DNW30 and DNW25 pixel designs (additional deep n-implantation with a size of \num{30} and \SI{25}{\micro\meter}). For the sake of clarity, other pixel designs are omitted here since they have a lower efficiency after irradiation.
\begin{figure}
	\begin{center}
		\includegraphics[width=0.8\linewidth]{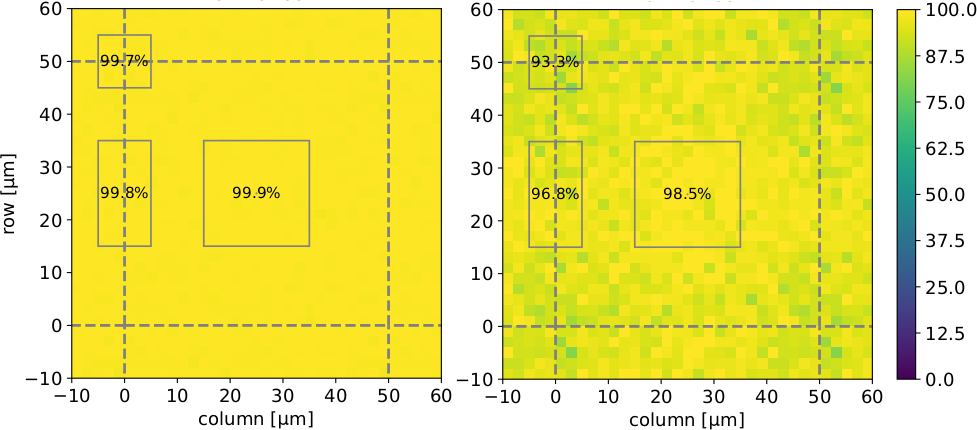}
		\caption{In-pixel efficiency map after a fluence of \SI{1e16}{\neq\per\square\centi\meter} for two different pixel flavors (left: NW30 and right: NW15) at a bias voltage of \SI{400}{\volt}.}
		\label{fig:in_pixel_eff}
	\end{center}
\end{figure}

Before irradiation, an efficiency larger than \SI{99.5}{\percent} is achieved at \SI{5}{\volt} only and there is no significant difference across the various pixel designs visible. After irradiation, the efficiency increases with increasing bias voltage. The measured efficiency after a fluence of \SI{5e15}{\neq\per\square\centi\meter} and a bias voltage of \SI{350}{\volt} is \SI[separate-uncertainty=true]{99.89(1)}{\percent} for the NW30 flavor and \SI[separate-uncertainty=true]{99.88(1)}{\percent} for the DNW30 flavor. This is well above the requirement of \SI{97}{\percent} after irradiation (grey dashed line in Fig.~\ref{fig:eff_vs_bias}). After a fluence of \SI{1e16}{\neq\per\square\centi\meter} the efficiency decreases further (especially for low bias voltages). However, at \SI{400}{\volt} an efficiency of \SI[separate-uncertainty=true]{99.79(1)}{\percent} for the NW30 and \SI[separate-uncertainty=true]{99.78(1)}{\percent} for the DNW30 flavor can still be achieved.
Especially, at the highest measured fluence, it is visible that flavors with a smaller implant size show a slightly lower efficiency\footnote{This is also true for the omitted (D)NW20 and (D)NW15 flavors.}. Furthermore, at low bias voltages, the hit-detection efficiency of designs with a deep n-well (same implant size) is higher compared to designs with only the standard n-well geometry, especially for smaller implant sizes.

Fig.~\ref{fig:in_pixel_eff} shows in-pixel efficiency maps (all data mapped onto a single pixel) at a fluence of \SI{1e16}{\neq\per\square\centi\meter} for two different pixel flavors (NW30 and NW15) at a bias voltage of \SI{400}{\volt}. One can see that the efficiency, for pixel designs with smaller implant size, is low at the pixel corners which is due to the low electric field (and charge sharing) in this region.

\section{Charge measurements and charge-collection efficiency}
In addition to the hit-detection efficiency, the charge collection behaviour was studied during test beams.
The readout chip already provides an internal charge information, called ToT (time over threshold). However, the precision of this measurement needed for a charge calibration (using radioactive sources) or charge measurements during test beams is not sufficient. This problem is circumvented by using the so-called \textit{TDC method} \cite{pohl_phd}. This method makes use of the chip's HitOR signal (logical OR of all discriminator outputs) whose length is proportional to the collected charge. This signal is sampled with a \SI{640}{\mega\hertz} clock provided externally by the readout system, thus enabling a precise charge measurement.
\begin{figure}
	\begin{center}
		\includegraphics[width=0.8\linewidth]{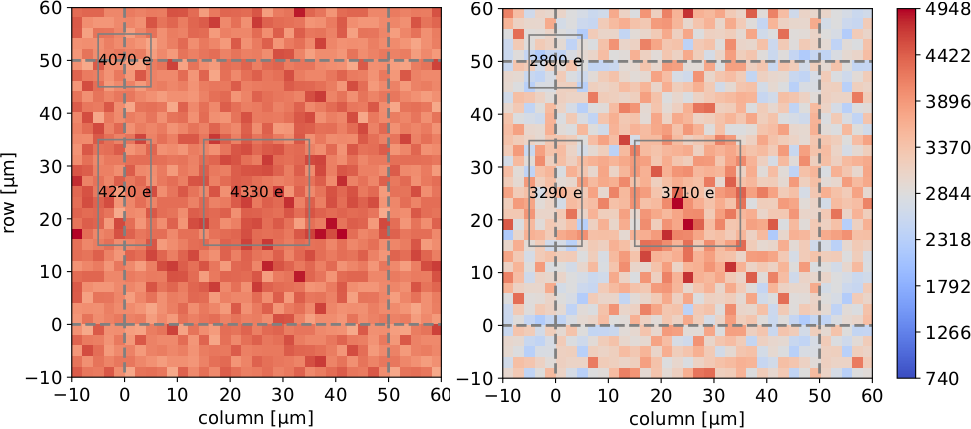}
		\caption{In-pixel charge map in electrons after a fluence of \SI{1e16}{\neq\per\square\centi\meter} for two different pixel flavors (left: NW30 and right: NW15) at a bias voltage of \SI{400}{\volt}.}
		\label{fig:in_pixel_charge}
	\end{center}
\end{figure}

In Sec.~\ref{sec:eff} it was shown that, after irradiation, the efficiency for pixel designs with smaller implant sizes is low in the pixel corners where the electrical field is low. The same behaviour (low charge in pixel corners) is observed for the collected charge. The corresponding in-pixel charge maps (only events with cluster size of 1), after a fluence of \SI{1e16}{\neq\per\square\centi\meter}, can be seen in Fig.~\ref{fig:in_pixel_charge}. This is in agreement with the expectations and explains the efficiency loss since a lower electric field after irradiation leads to more charge carrier trapping, and thus to a smaller collected charge which in turn results in a lower efficiency for a given threshold.

\begin{figure}
	\begin{center}
		\includegraphics[width=0.7\linewidth]{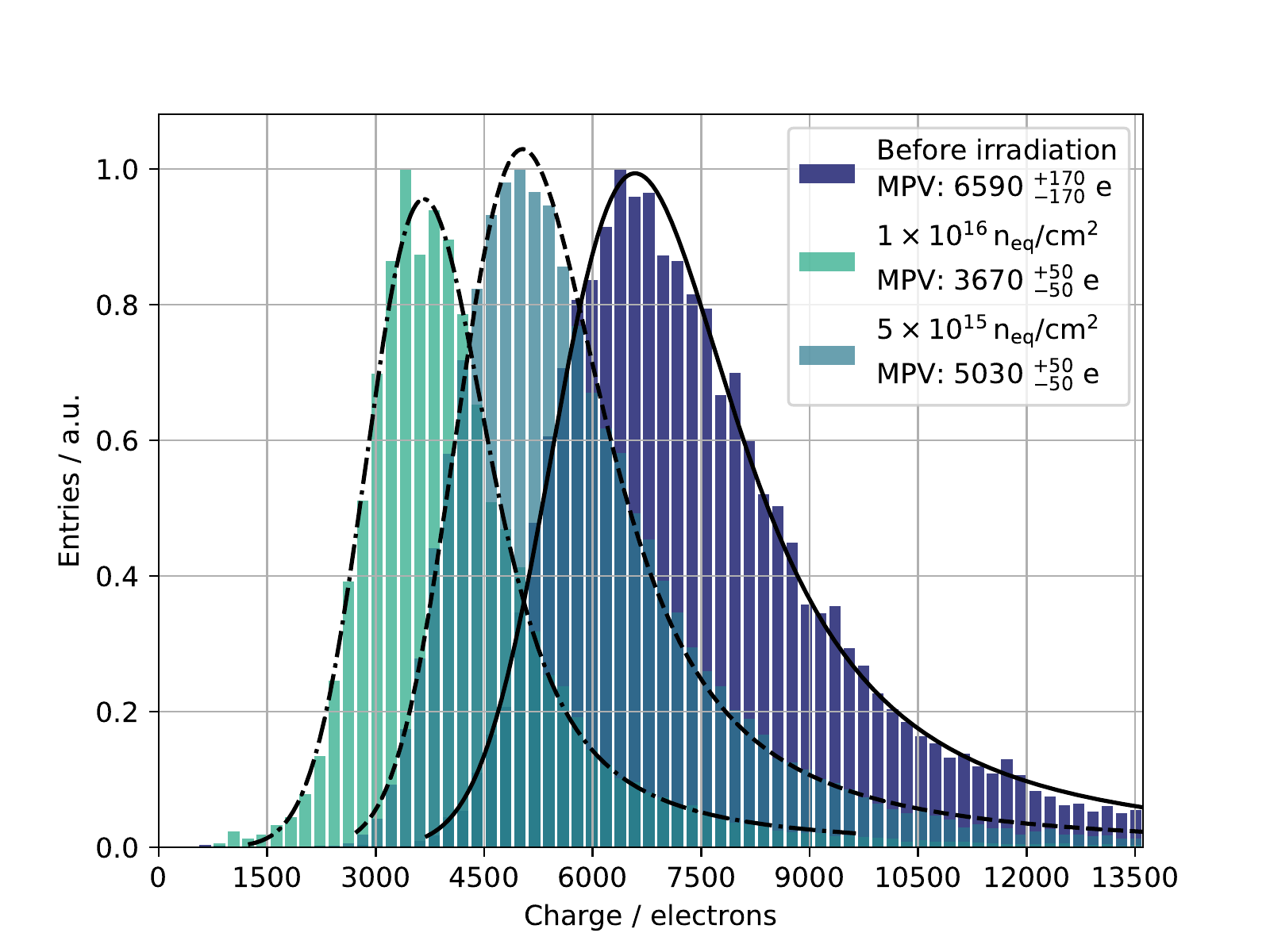}
		\caption{Measured charge distributions for different irradiation levels for the NW30 flavor. The distributions are fitted with a Landau-Gauss convolution to extract the most probable value (MPV). Before irradiation a MPV of \SI[separate-uncertainty=true]{6590(170)}{\electrons} is measured at \SI{80}{\volt}. At a fluence of \SI{5e15}{\neq\per\square\centi\meter} a MPV of \SI[separate-uncertainty=true]{5030(50)}{\electrons} is measured at \SI{350}{\volt}, and at a fluence of \SI{1e16}{\neq\per\square\centi\meter} a MPV of \SI[separate-uncertainty=true]{3670(50)}{\electrons} is measured at \SI{400}{\volt}.}
		\label{fig:charge_spectra}
	\end{center}
\end{figure}

 In Fig.~\ref{fig:charge_spectra} the measured charge distributions for the different irradiation levels measured using the NW30 pixel design are depicted. The distributions follow a Landau function convoluted with a Gaussian function due to electronic noise. Furthermore, it is visible that the most probable value (MPV) extracted from a fit decreases with increasing fluence. The reasons for that are the facts that a) after irradiation the sensor can no longer be fully depleted at reasonable bias voltages and b) charge carrier trapping during charge col\-lection due to radiation damage of the bulk.

\begin{figure}
	\begin{center}
		\includegraphics[width=0.7\linewidth]{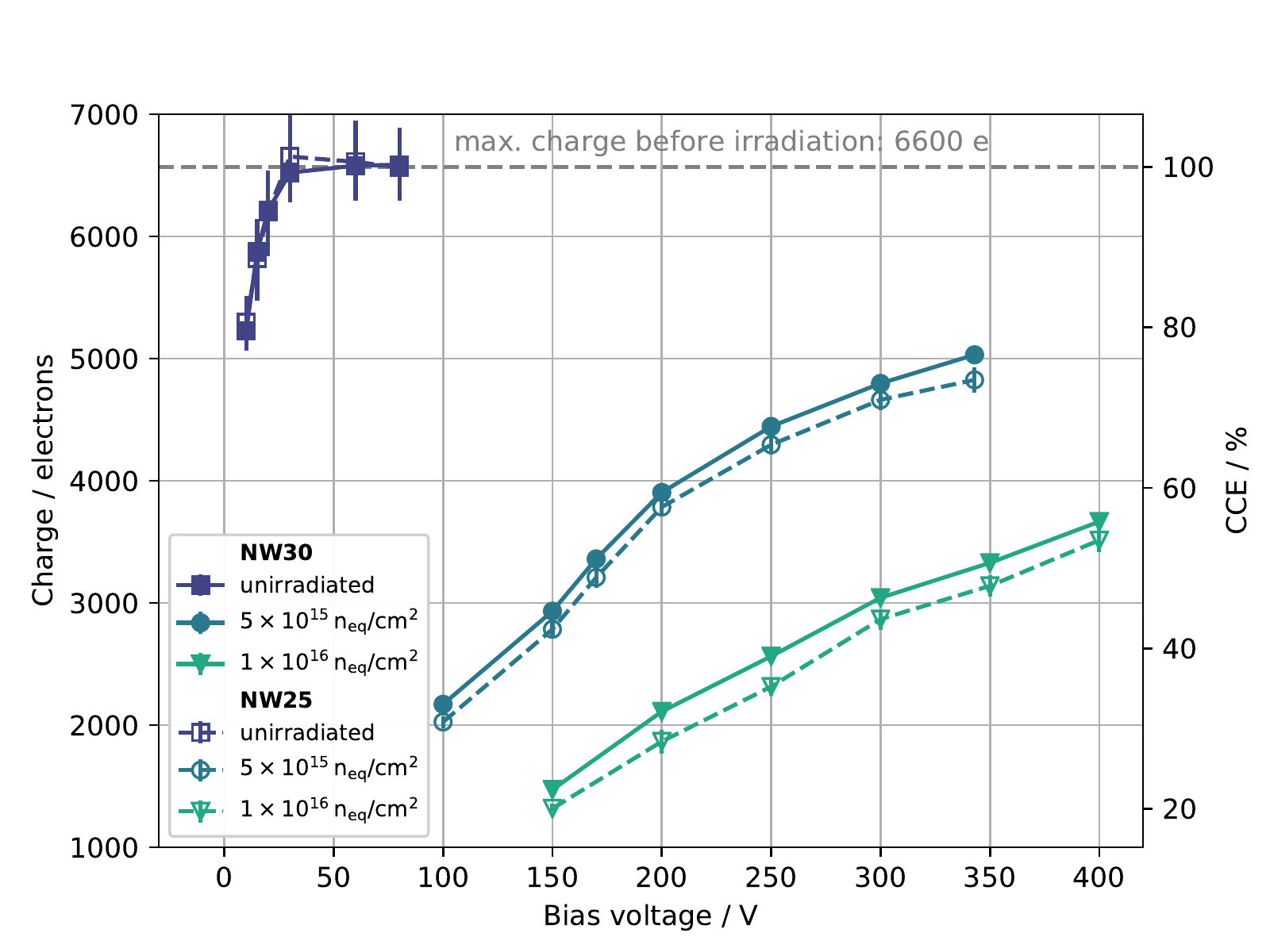}
		\caption{Collected charge as a function of the bias voltage for different irradiation levels. Data points are the most probable values extracted from a fit to the measured charge distributions. The error bars originate from the fit. The y-axis on the right-hand side shows the charge collection efficiency (CCE).}
		\label{fig:charge_vs_bias}
	\end{center}
\end{figure}

The collected charge as a function of the bias voltage, illustrated in Fig.~\ref{fig:charge_vs_bias}, was studied. Each data point corresponds to the most probable value extracted from a fit to the measured charge distribution. The charge signal increases with bias voltage since the depleted volume extends with increasing bias voltage. Before irradiation, the amount of collected charge starts to saturate at approximately \SI{40}{\volt} leading to a charge signal of about \SI{6600}{\electrons}. This indicates that the sensor is fully depleted at approximately \SI{40}{\volt}. Assuming that \num{73} electrons per \si{\micro\meter} (extracted from a GEANT4 simulation) are created within the depletion zone, this yields a silicon bulk thickness of approximately \SI{90}{\micro\meter}. This values is reasonable since the nominal thickness of \SI{100}{\micro\meter} includes also the metal layers with a thickness of a few \si{\micro\meter}. After irradiation, the amount of collected charge decreases due to the fact that the sensor can no longer be fully depleted and charge carrier trapping sets in. The measured charge signal after a fluence of \SI{1e16}{\neq\per\square\centi\meter} is approximately \SI{3700}{\electrons} at the highest measured bias voltage of \SI{400}{\volt}.

The measured charge signal can be translated to a charge collection effi\-ciency (CCE), as shown in Fig.~\ref{fig:charge_vs_bias} (axis on the right-hand side). The charge collection efficiency is obtained by dividing the measured charge by the maximum measured charge before irradiation (\SI{6600}{\electrons}). This yields a maximum charge collection efficiency of approximately \SI{80}{\percent} at a fluence of \SI{5e15}{\neq\per\square\centi\meter} and \SI{55}{\percent} at a fluence of \SI{1e16}{\neq\per\square\centi\meter}, respectively.

\section{Conclusion and outlook}
The radiation tolerance of $\SI{100}{\micro\meter}$ thin passive CMOS sensors fabricated in \SI{150}{\nano\meter} LFoundry technology has been investigated. The sensors are still functional even after a fluence of \SI{1e16}{\neq\per\square\centi\meter} and can be successfully operated. At this fluence a hit-detection efficiency of \SI[separate-uncertainty=true]{99.79(1)}{\percent} is measured (at \SI{400}{\volt}) for the NW30 design. The charge collection efficiency is measured to be approximately \SI{55}{\percent} at the maximum tested bias voltage of \SI{400}{\volt} after the highest fluence. In addition, the power dissipation of the sensor needed to meet the ATLAS ITk requirements in terms of efficiency is comparable to 3D sensors.
This demonstrates that passive CMOS sensors are radiation tolerant and withstand a fluence of \SI{1e16}{\neq\per\square\centi\meter}, the expected fluence for the future innermost ATLAS pixel detector layer.
The performance of passive CMOS sensors in terms of noise and hit-detection efficiency equals that of conventional planar sensors.
Full-size (RD53-sized) passive CMOS sensors using the NW30 geometry for the ATLAS and CMS experiments at the HL-LHC are already manufactured and currently under investigation.

\section*{Acknowledgements}
We would like to thank LFoundry and Ion Beam Services (IBS) for the fabrication and the processing of the backside of the sensors.
We also thank Laura Gonella for making the irradiation at the Birmingham Irradiation Facility possible. Further, we would like to thank the HISKP group for making the irradiation at the Proton Irradiation Site in Bonn possible.
This project has received funding from the Deutsche Forschungsgemeinschaft DFG, under grant agreement no. WE 976/4-1, the German Ministerium f\"ur Bildung, Wissenschaft, Forschung und Technologie (BMBF) under contract no. 05H15PDCA9, the H2020 project AIDA-2020, under grant agreement no. 654168, and from a Marie Sklodowska-Curie ITN Fellowship of the European Union’s Horizon 2020 program under grant agreement no. 675587-STREAM.
The measurements leading to these results have been performed at the Test Beam Facility at DESY Hamburg (Germany), a member of the Helmholtz Association (HGF)

\bibliography{mybibfile}

\end{document}